\documentclass[prl,twocolumn,showpacs,preprintnumbers,superscriptaddress,
amsmath]{revtex4}

\usepackage{tikz}
\usetikzlibrary{arrows,positioning}
\usepackage{verbatim}
\usepackage{color}
\usepackage{times}
\usepackage{subfigure}
\usepackage{dcolumn}
\usepackage{bm}
\usepackage{multirow}

\newcommand{\bra}[1]{\mathinner{\langle{#1}|}}
\newcommand{\ket}[1]{\mathinner{|{#1}\rangle}}

\include{partition}

\begin{document}

 \title{Rapid and robust spin state amplification}

 \author{Tom Close$^\dagger$}
 \email{tom.close@materials.ox.ac.uk}
 \affiliation{Department of Materials, Oxford University, Oxford OX1 3PH, UK}
 \author{Femi Fadugba\footnote{These authors have contributed equally to the work reported here.}}
 \affiliation{Department of Materials, Oxford University, Oxford OX1 3PH, UK}
 \author{Simon C. Benjamin}
 \affiliation{Department of Materials, Oxford University, Oxford OX1
   3PH, UK}
 \affiliation{Centre for Quantum Technologies, National University of Singapore, 3 Science Drive 2, Singapore 117543}
 \author{Joseph Fitzsimons}
 \affiliation{Centre for Quantum Technologies, National University of Singapore, 3 Science Drive 2, Singapore 117543}
 \author{Brendon W. Lovett}
 \affiliation{Department of Materials, Oxford University, Oxford OX1 3PH, UK}
 \affiliation{School of Engineering and Physical Sciences, Heriot-Watt University, Edinburgh EH14 4AS}

 \begin{abstract}
   Electron and nuclear spins have been employed in many of the early
   demonstrations of quantum technology (QT). However applications in
   real world QT are limited by the difficulty of measuring single
   spins. Here we show that it is possible to rapidly and robustly
   amplify a spin state using a lattice of ancillary spins. The model
   we employ corresponds to an extremely simple experimental system:
   a homogenous Ising-coupled spin lattice in one, two or three
   dimensions, driven by a continuous microwave field. We establish
   that the process can operate at finite temperature (imperfect
   initial polarisation) and under the effects of various forms of
   decoherence.
 \end{abstract}

 \maketitle

 The standard approach to implementing a quantum technology is to
 identify a physical system that can represent a qubit: it must exhibit
 two (or more) stable states, it should be manipulable through external
 fields and possess a long decoherence time. Provided that the system
 can controllably interact with other such systems, then it may be a
 strong candidate. Electron and nuclear spins, within suitable
 molecules or solid state structures, can meet these
 requirements. However the drawback with spin qubits is that they have
 not been directly measured through a detection of the magnetic field
 they produce. The magnetic moment of a single electron spin is orders
 of magnitude too weak to be detected by standard ESR techniques and
 even the most sensitive magnetometers still fall short of single spin
 measurement~\cite{cleuziou06} - meanwhile the situation with nuclear spins is
 worse still. In a few special systems it is possible to convert the
 spin information into another degree of freedom. For example, a
 spin-dependent optical transition allow spin to photon conversion in
 some crystal defects~\cite{jelezko04, neumann10, Morello:2010p6617},
 self-assembled semiconductor quantum dots~\cite{vamivakas10, Berezovsky:2008p6616},
 and
 trapped atoms held in a vacuum~\cite{leibfried03}. Alternatively, spin
 to charge conversion is an established technology in lithographic
 quantum dots~\cite{hanson07}. However, the majority of otherwise
 promising spin systems do not have such a convenient
 property~\cite{morton10} and therefore cannot be measured directly.

 One suggested solution is to `amplify' a single spin, by using a set
 of ancillary spins that are (ideally) initialised to $\ket{0}$. We would look for a transformation of the form
 \begin{equation}
 \ket{0}\ket{0}^{\otimes n} \rightarrow \ket{0}\ket{0}^{\otimes n} \ \ \ \ \ \
 \ket{1}\ket{0}^{\otimes n} \rightarrow \ket{1}\ket{1}^{\otimes n},
 \end{equation}
 the idea being that the $n$ ancillary spins constitute a large enough
 set that state of the art
 magnetic field sensing technologies can detect them. Note that the transformation need not be unitary or indeed even coherent: since the intention is to make a measurement of the primary spin, it is not necessary to preserve any superposition (that is, we need not limit ourselves to transformations that take $\alpha\ket{0}\ket{0}^{\otimes n} +\beta\ket{1}\ket{0}^{\otimes n} $ to a cat state like $\alpha\ket{0}^{\otimes n+1}+\beta\ket{1}^{\otimes n+1}$).

 This is a rather broadly defined transformation and there are a number
 of ways that one might perform it. Clearly one would like to find the
 method that is the least demanding experimentally. Previous authors
 have proposed schemes using a strictly one-dimensional (1D)
 homogeneous lattice with continuous global driving
 \cite{Lee:2005p6468}, and an inhomogeneous three-dimensional (3D)
 lattice with alternating timed EM pulses
 \cite{PerezDelgado:2006p6542}. The former result has the advantage of
 simplicity but the rate at which amplification occurs will inevitably
 be limited by the single dimension of the array; moreover such a
 system must be highly vulnerable to imperfect initialisation
 (i.e. finite temperature). Here we generalise to a homogeneous
 two-dimensional (2D) square lattice, showing that a continuous global
 EM field can drive an amplification process that succeeds at finite
 temperatures (imperfect initialisation of the ancilla spins) and in
 the presence of decoherence. By bringing the global EM field onto
 resonance with certain transitions, we are able to create a set of
 rules that govern locally how spins propagate over the lattice. We
 then look at the rate of increase in the total number of flipped spins
 as a measure of quality of the scheme. While our focus is on the 2D
 case, we are also able to predict the performance of the amplification
 protocol for a homogeneous 3D lattice with continuous driving.

 The case of a 1D lattice has been studied in detail by
 Lee and Khitrin \cite{Lee:2005p6468}. Before moving to the 2D
 spin lattice that will form the core of the paper, we first recall how to simplify the description of this (semi-infinite) 1D spin chain, with nearest neighbour Ising (ZZ)
 interactions. Under a microwave driving field of frequency $\omega$,
 the Hamiltonian is given by
 \begin{equation} {\cal H} = \sum_{i=1}^{\infty} \epsilon_i \sigma^i_z
   + J_i \sigma_z^i\sigma_z^{i+1} + 2 \Omega_i \sigma_x^i \cos(\omega
   t)
 \end{equation}
 $\epsilon_i$ is the on-site Zeeman energy of spin $i$, and $J_i$ is
 the magnitude of the coupling between spins $i$ and $i+1$. $\Omega$
 describes the coupling of spin $i$ and the microwave field. In this case, spin $i=1$ is the one whose state is supposed to be amplified. If we assume that the chain is uniform, such that
 $\Omega_i = \Omega$, $\epsilon_i = \epsilon$ and $J_i = J$, then moving
 into a frame rotating at frequency $\omega$, making a rotating wave
 approximation and setting $\omega = \epsilon$ leads to
 \begin{equation} {\cal H} = \sum_{i=1}^{\infty} J
   \sigma_z^i\sigma_z^{i+1} + \Omega \sigma_x^i .
 \end{equation}
 In order to understand the dynamics of the system, is it instructive to
 explicitly separate all terms that involve a particular spin $k$:
 \begin{equation} {\cal H} = J (\sigma_z^{k-1} +
   \sigma_z^{k+1})\sigma_z^k + \Omega \sigma^k_x + \sum_{i \neq \{k,
     k-1\}} \Omega\sigma_x^i + J \sigma_z^i\sigma_z^{i+1} +
   \Omega\sigma_x^{k-1}
 \label{hamk}
 \end{equation}
 Choosing a driving field such that $\Omega\ll J$ means that
 spin $k$ will only undergo resonant oscillations when the first term
 in Eq.~\ref{hamk} goes to zero - i.e. when the two spins neighbouring
 spin $k$ are oriented in opposite directions. In any other
 configuration the Ising coupling takes the spin $k$ off resonance with
 the microwave and no appreciable dynamics are expected.

 Let us now define a subset of states $S$ that exist in the spin chain
 Hilbert space, $\ket{n}$, which have the first $n$ spins of the chain
 in state $\ket{\uparrow}$ with the rest $\ket{\downarrow}$. If the
 rule we just derived holds exactly these states define a closed
 subspace. We may then write a very simple isolated Hamiltonian
 for this subspace:
 \begin{equation}
 {\cal H}_S = \Omega \sum_{n=1}^\infty \ket{n}\bra{ n+1}+\ket{n+1}\bra{ n}.
 \label{1d_ham}
 \end{equation}

 With this simplification of the 1D Hamiltonian in mind, we progress now to a semi-infinite square spin lattice with nearest-neighbour ZZ interactions. For this case we have
 \begin{equation} {\cal H} = \sum_{i=1}^{\infty}\sum_{j=1}^\infty
   \epsilon \sigma^{i, j}_z + J \sigma_z^{i, j}\sigma_z^{i+1, j} +
   J\sigma_z^{i, j}\sigma_z^{i, j+1} + 2 \Omega \sigma_x^{i, j}
   \cos(\omega t).
 \end{equation}
 By again considering the terms affecting a particular spin in the main
 body of the lattice ($k(>1), l(>1)$ say) we find for $\omega =
 \epsilon$ and after moving to a rotating frame and making the rotating
 wave approximation:
 \begin{equation}
 {\cal H} = J \sigma_z^{k, l} (\sigma_z^{k+1, l} + \sigma_z^{k, l+1} + \sigma_z^{k-1, l} + \sigma_z^{k, l-1}) +...
 \end{equation}
 where we do not explicitly write out terms not involving spin $(k, l)$.
 The microwave is now only resonant for spin $(k, l)$ if it has two
 neighbour spins in each orientation. For a spin on the edge of the
 lattice there are an odd number of neighbours so resonance cannot be
 achieved in this case. However, applying a second microwave with
 $\omega = \epsilon - J$ allows resonant flips on the edge if two
 neighbours are down and one up - and this second field has no effect on the
 bulk spins.

 The spin to be measured is the corner spin ($i=j=1$) and so would form part of a wider computational apparatus. We may therefore assume that it is a different species with
 a unique resonant frequency. The dynamics of the whole lattice may then be summarised by three rules (in order of precedence):
 \begin{enumerate}
   \item The corner (test) spin is fixed.\vspace{-0.2cm}
   \item An edge spin can flip if it has one of its neighbours up and
     two down.\vspace{-0.2cm}
   \item A body spin can flip if it has two of its neighbours up and two down.\vspace{-0.2cm}
 \end{enumerate}
 We begin by supposing all spins are initialised in the `down' state
 apart from the test spin, which is located in the upper left hand corner
 of our lattice. We can describe this initial state by choosing two
 basis elements: $\ket 0$ when the test spin is down, and $\ket 1$
 when the test spin is up. Using our heuristic rules we can see that these two states do not
 couple to each other - that $\bra 0H\ket 1=0$. In fact $\ket 0$
 does not couple to any other state, so if we start in the $\ket 0$
 state no amplification occurs, as desired.

 We will now seek to construct a basis for the subspace containing our
 system evolution, by looking at states connected by our
 Hamiltonian. It will be convenient to represent these states on the nodes of
 a graph, using the edges to represent non-zero elements of the
 Hamiltonian.

\begin{figure}
     \includegraphics[scale=0.6]{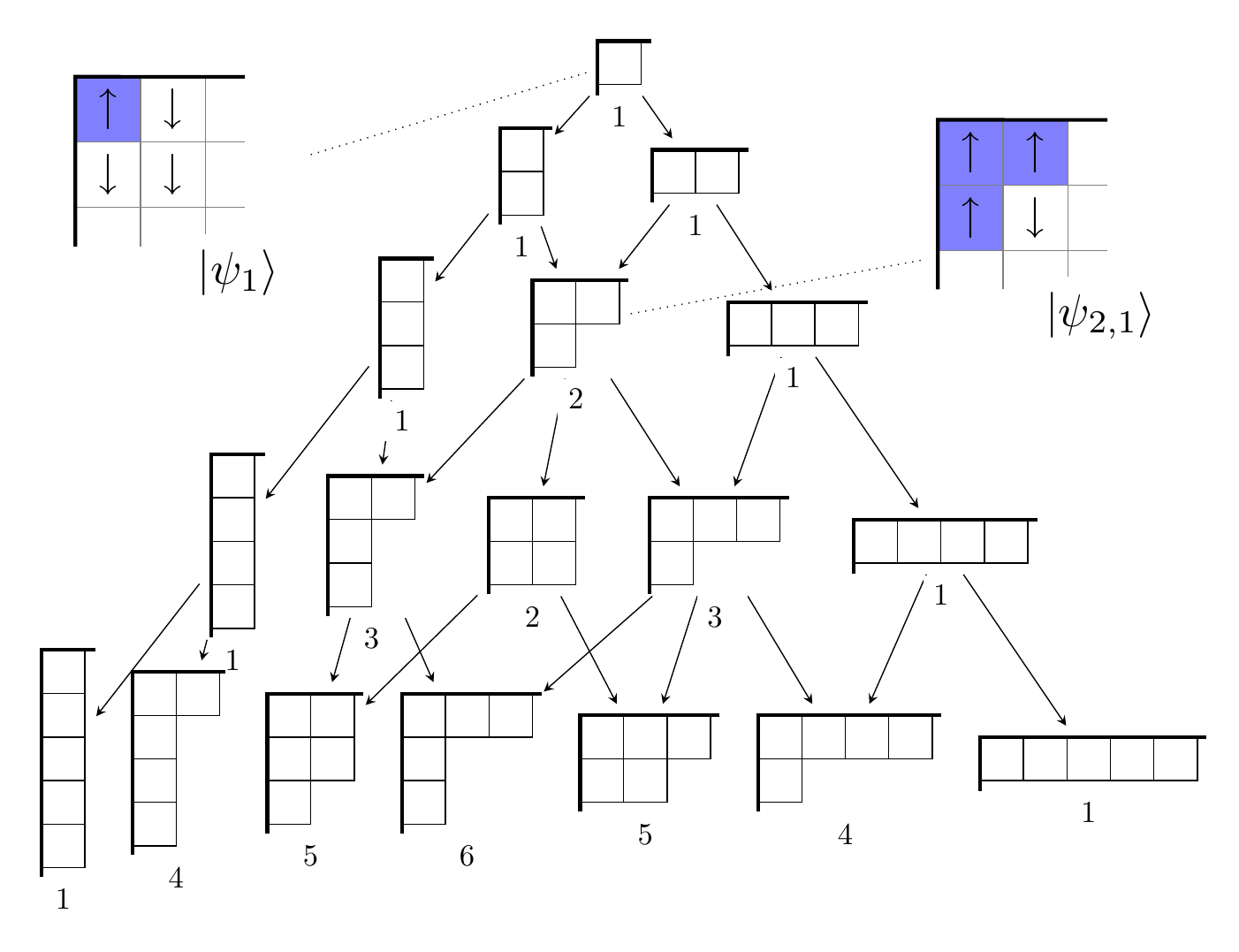}
   \caption{Partition states arranged into a lattice. Edges represent a coupling
   through the Hamiltion of strength $\Omega$. Weights represent the
   number of different paths through the lattice to a given state.}
   \label{partition_states}
 \end{figure}

 Our starting point is the state $\ket 1$, with just the corner spin
 `up'. From this position our rules allow two possibilities: either the
 spin to the right of the corner flips, or the spin below it flips (see Fig.~\ref{partition_states}). In each case the
 magnitude of the transition matrix element is $\Omega$. As we continue this procedure, we notice that the states that arise for each excitation number
 can be characterised by a non-increasing sequence of
 integers that represent the number of `up'-spins in each column of the
 lattice (see Fig.~\ref{partition_states}). Such sequences can also be used to define partitions of at
 integer: ways of splitting an integer up into a sum of other integers,
 e.g. $3=3=2+1=1+1+1$. In fact, the states that arise are in 1-to-1
 correspondence with such partitions; we call these states
 `partition states' and denote them with standard partition notation
 (see Fig.~\ref{partition_states}).
 The graph we have just described is depicted in
 Fig.~\ref{partition_states}  is known as `Young's lattice' and arises
 in areas of pure mathematics, such as the representation theory of the
 symmetric group, and the theory of differential posets. We have drawn
 weights beneath each state, recording the number of ways the state can
 be constructed. We will now further reduce the dimension of this basis
 by eliminating combinations of states which are inaccessible.

 Starting with $\ket 1$ we see that $\bra
 1H\left(\alpha_{1,1}\ket{\psi_{1,1}}+ \alpha_{2}\ket{\psi_{2}}\right)
 =\Omega\left(\alpha_{1,1}+ \alpha_{2}\right)$ so $\ket 1$ does not
 couple to the two-excitation state $\ket{\psi_{1,1}}-\ket{\psi_{2}}$. We can
 eliminate this, leaving a single orthogonal, coupled state with two excitations: $\ket 2 :=
 \frac{1}{\sqrt{2}}\left( \ket{\psi_{1,1}}+\ket{\psi_2} \right)$.

 We may continue to build up coupled states with larger excitation numbers, and in fact we find that there is only a single coupled state in each case (i.e. we can always eliminate $k-1$ combinations of partition states with $k$ excitations).
 To see this, first suppose we have the coupled state with $k$
 excitations, which by analogy with the 1D case we write as $\ket
 k$. We can write $\ket k=\frac{1}{N_k}\sum_{i\in
   P(k)}c_{i}\ket{\psi_{i}}$, where $P(k)$ is the set of partitions of
 the integer $k$ and $N_k$ a normalisation factor. We want to construct
 the state $\ket{k+1}$ by eliminating the $k$-dimensional subspace with
 $k+1$ excitations, to which $\ket{k}$ does not couple.

 Let $\ket{\psi}=\sum_{j\in P(k+1)}\alpha_{j}\ket{\psi_{j}}$ and
 consider the states $\ket{\psi}$ such that
 \[
 0=\bra kH\ket{\psi}=\sum_{i\in P(k)}\sum_{j\in P(k+1)}c_{i}^{*}\alpha_{j}\bra{\psi_{i}}H\ket{\psi_{j}}\]
 but $\bra{\psi_{i}}H\ket{\psi_{j}}=\Omega$ if $i$ is a {\it parent} of $j$
 (a state connect to $j$, in the lattice row above it), and $0$ otherwise, so
 \[0=\bra kH\ket{\psi}=\sum_{j\in P(k+1)}\alpha_{j}\sum_{i\in parents(j)}c_{i}^{*}.\]
 This is the equation of a hyperplane in $|P(k+1)|$ dimensions,
 defining the states that are not coupled to $\ket k$ through the Hamiltonian.
 There is a unique single state orthogonal to this hyperplane,
 $\beta_{j}=\sum_{i\in parents(j)}c_{i}$, to which $\ket k$ couples.
 So the only state with $k+1$ `up'-spins that $\ket k$ couples has coefficients
 proportional to $\beta_{j}$. After normalisation, we call this state
 $\ket{k+1}$.

 Unfortunately there is no easy way to write down the partition states
 and weights for the $n$th row of the lattice. Fortunately, for our purposes, we only need to know
 that the states $\ket k$ exist and what the coupling between them
 is. To find this coupling, consider
 \begin{align}
 g_{n-1,n} & = \bra nH\ket{n-1} \nonumber\\
  & = \frac{1}{N_{n-1}N_n}\sum_{i\in P(n)}\sum_{j\in
    P(n-1)}c_{i}^{*}c_{j}\bra{\psi_{i}}H\ket{\psi_{j}} \nonumber\\
  & = \frac{1}{N_{n-1}N_n}\Omega\sum_{i\in P(n)}c_{i}^{*}\sum_{j\in
    parents(i)}c_{j} \nonumber\\
  & = \frac{1}{N_{n-1}N_n}\Omega\sum_{i\in P(n)}|c_{i}|^{2} =
  \Omega\frac{ N_n}{N_{n-1}}\label{coupling_as_sum_c2}\end{align}
 To find the $N_n$ we need the sum of the squares of the weights of partitions
 in a given row. A standard result about Young's lattice
  immediately gives us this sum: $n!$ \cite{STANLEY:1975p6605}.

 Referring back to Eq. (\ref{coupling_as_sum_c2}), and using
  $N_i=\sqrt{i!}$, we see that
 \begin{equation}
 {\cal{H}} = \Omega \sum_{n} \sqrt{n} \left( \ket{n-1} \bra{n} +
 \ket{n} \bra{n-1} \right) .
 \label{2d_ham}
 \end{equation}
 In essence we
  have established a linear sequence of states, each coupled to the
  the next analogously to the states on a 1D chain \ref{1d_ham}. However, each of our states
  is in fact a superposition of many configurations of the 2D array,
  and crucially the effective coupling
  from each state to the next increases along the sequence.


 It has been shown (e.g. \cite{Fitzsimons:2005p6472}) that a quantum state released at
 the end of a semi-inifinite chain of states, with constant couplings, will
 travel ballistically: the average position of the state along the
 chain is
 proportional to the time passed, and inversely proportional to the
 coupling strength. Since, in the one-dimensional case, the position is
 proportional to the number of spins that have flipped, we have that
 the total polarisation will increase linearly with time.

\begin{figure}
 \includegraphics[scale=0.5]{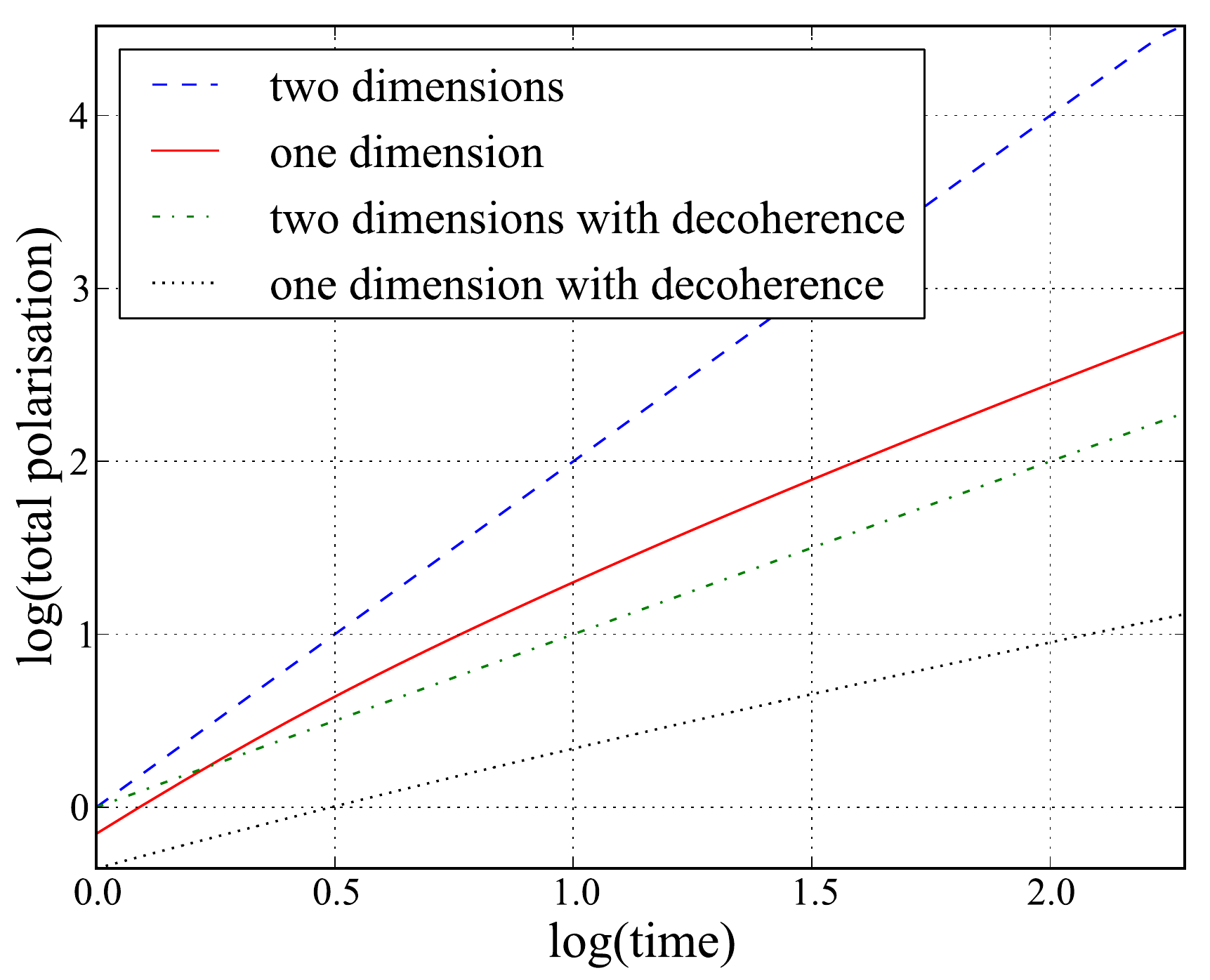}
 \caption{Expected total polarisation against time. Time in units of
   $\frac{1}{\Omega}$, dephasing rate $\Gamma = 1$. The gradient of the
 'one dimension with decoherence' line tends to $\frac{1}{2}$ asymptotically.}
 \label{comparison}
 \end{figure}

 We can establish the rate of propagation in the 2D case using the ansatz
 that the time taken to travel between two neighbouring nodes is inversely
 proportional to the strength of the coupling between them. The total
 time is then $t_{2D}\propto \sum_{i=1}^{n}\frac{1}{\sqrt{i}}\simeq
 n^{\frac{1}{2}}$. As in the one-dimensional case, the position along the chain
 corresponds to the the number of spins that have flipped, and so we
 would expect the total polarisation to be proportional to $t^2$. This prediction of a quadratic speed-up of signal going from 1D to 2D is the central result of our paper, and
 was confirmed by simple numerical simulations of Eq. (\ref{2d_ham})
 (Fig. \ref{comparison}).



 Unfortunately the mapping from 2D to 1D is not readily extendible to 3D. However,
 our results so far could have been anticipated using simple dimensional arguments; if one postulates that the rate of spin propagation is
 proportional to the boundary of the region, one can predict the
 correct scaling behaviour. In 1D the boundary size is
 independent of the region size; no matter how many spins have flipped,
 it still has size one. The coupling strength between states $\ket{n}$
 is constant. In the 2D case, the boundary size scales
 with the square root of the area, and the coupling goes with
 $\sqrt{n}$. In 3D, the boundary scales like the cube root of the volume squared, and so we expect the coupling to scale
 as $n^{\frac{2}{3}}$. Following similar logic to that used in 2D
 case: $t_{3D}\propto \sum_{i=1}^{n}\frac{1}{i^\frac{2}{3}}\simeq
 n^{\frac{1}{3}}$, and so $n \sim t^3$.


We now consider the effect of decoherence. Much of the early work on
continuous time quantum random walks looked at the speedup they afforded over their classical
 counterparts~\cite{Farhi:1998p6471}, but didn't make  any statement about the conditions
 under which we would expect the quantum walk to exhibit classical
 behaviour, as we might expect in a regime of suitably heavy
 dephasing, say.

We begin by considering a collective noise operator: $
L=\sum_{n}n\ket n\bra n$.
This represents noise that applies uniformly to the whole lattice:
global fluctuations in the magnetic field, for example. As the effect
of this type of noise depends only on the number of `up' spins, the
system remains in the reduced basis of number states calculated earlier,
with only the coherences between these states affected.

Our starting point is the Lindblad master equation
\begin{equation}
\dot{\rho}=i\left[\rho,H\right]+\frac{1}{2}\Gamma\left(2L\rho
  L^{\dagger}-L^{\dagger}L\rho-\rho L^{\dagger}L\right).
\end{equation}
 We proceed by splitting up the equation into diagonal and off-diagonal
terms:\begin{align} \dot{\rho}_{ii} & = i\sum_{k=\pm
    i}\left(\rho_{ik}g_{ki}-\rho_{ki}g_{ik}\right)=-2\sum_{k=\pm
    i}Re\left[\rho_{ik}g_{ki}\right] \label{rhoii}\\
  \dot{\rho_{ij}} & = i\left(\sum_{k=\pm j}\rho_{ik}g_{kj}-\sum_{k=\pm
      i}\rho_{kj}g_{ik}\right)-\Gamma\rho_{ij}\end{align} where
$g_{ij}$ is the coupling between states $i$ and $j$. In the limit of
heavy dephasing ($\Gamma\gg g$), we have a process similar to
adiabatic following, and we can make the approximation\[
\Gamma\rho_{ij}\approx i\left(\sum_{k=\pm
    j}\rho_{ik}g_{kj}-\sum_{k=\pm i}\rho_{kj}g_{ik}\right).\] We
consider the $\rho_{ij}$ as a set of $\frac{n(n-1)}{2}$ variables and
solve for them in terms of the $\rho_{ii}$. Neglecting terms that are
second order in $\frac{g}{\Gamma}$, and substituting back into
Eq. (\ref{rhoii}) gives\[
\dot{\rho}_{ii}=-\sum_{j=i\pm1}\frac{2|g_{ij}|^{2}}{\Gamma}\left(\rho_{ii}-\rho_{jj}\right).\]
Our quantum chain formally reduces to a classical Markov chain on the same
statespace, with transition rates proportional to the coupling
squared.

Although states with more `up' spins decohere more quickly, the
decoherence rate $\Gamma$ is not multiplied for higher states, as it is
the \textit{relative} decoherence rate between neighbouring states,
which is of importance.

In one-dimension $g_{ij} = 1$ and we are reduced to a simple random
walk on a semi-infinite line. By analogy with simple diffusion we
expect that the resulting distribution is
roughly Gaussian, with the expected number of flipped spins going with $\sqrt{t}$: the
rate of spin propagation drops from $t$ to $\sqrt{t}$. This result was
confirmed numerically (Fig. \ref{comparison}).

In the two-dimensional case $g_{ij} = \sqrt{j}, j=i+1$: We get a
random walk with increasing transition rates. Numerically (Fig. \ref{comparison}), we find that the rate of
spin propagation drops from $t^2$ to $t$ - still an encouraging
scaling.


We also investigated the `individual noise' case, where the dephasing
occurs independently on each site (see supplementary
material). Although the calculations differ, we see the same rate
behaviour as for the collective noise case.



Finally we consider imperfect inital polarisation (i.e. finite
temperature) - a property exhibited by any real experimental
system. As discussed in the supplementary material \footnote{Supplementary material: http://qunat.org/papers/amp/}, a fortuitous
consequence of the propagation rules is that our system is particularly
robust against this source of error; below an initialization threshold
of approximately $4\%$ it is extremely unlikely that a false positive
will occur. This places our protocol well within experimental
capabilities; for example for an array placed in a standard W-band electron spin resonance
system ($100$ GHz) and cooled using liquid 4He to $1.4$ degrees
Kelvin, only $3.1\%$ of electron spins will be in the `up' state.


\begin{acknowledgements}
We thank Gerard Milburn and John Morton for useful discussion. This work was supported by the EPSRC, the National Research Foundation and Ministry of Education, Singapore, and the Royal Society.
\end{acknowledgements}

\bibliography{spin_amp}

\end{document}